\documentclass[aps,floats,floatfix,superscriptaddress,twocolumn]{revtex4}
\usepackage{amsmath,amssymb}
\usepackage[pdftex]{graphicx}
\usepackage[tight]{subfigure}
\usepackage{color}

\begin{document}

\title{The Proof of Innocence}

\author{Dmitri Krioukov}
\affiliation{Cooperative Association for Internet Data Analysis
(CAIDA), University of California, San Diego (UCSD), La Jolla, CA
92093, USA}

\begin{abstract}

We show that if a car stops at a stop sign, an observer, e.g., a police officer, located at a certain distance perpendicular to the car trajectory, must have an illusion that the car does not stop, if the following three conditions are satisfied: (1)~the observer measures not the linear but angular speed of the car; (2)~the car decelerates and subsequently accelerates relatively fast; and (3)~there is a short-time obstruction of the observer's view of the car by an external object, e.g., another car, at the moment when both cars are near the stop sign.

\end{abstract}

\maketitle

\section{Introduction}

It is widely known that an observer measuring the speed of an object passing by, measures not its actual linear velocity by the angular one. For example, if we stay not far away from a railroad, watching a train approaching us from far away at a constant speed, we first perceive the train not moving at all, when it is really far, but when the train comes closer, it appears to us moving faster and faster, and when it actually passes us, its visual speed is maximized.

This observation is the first building block of our proof of innocence. To make this proof rigorous, we first consider the relationship between the linear and angular speeds of an object in the toy example where the object moves at a constant linear speed. We then proceed to analyzing a picture reflecting what really happened in the considered case, that is, the case where the linear speed of an object is not constant, but what is constant instead is the deceleration and subsequent acceleration of the object coming to a complete stop at a point located closest to the observer on the object's linear trajectory. Finally, in the last section, we consider what happens if at that critical moment the observer's view is briefly obstructed by another external object.

\section{Constant linear speed}

\begin{figure}
\centerline{\includegraphics[width=3.5in]{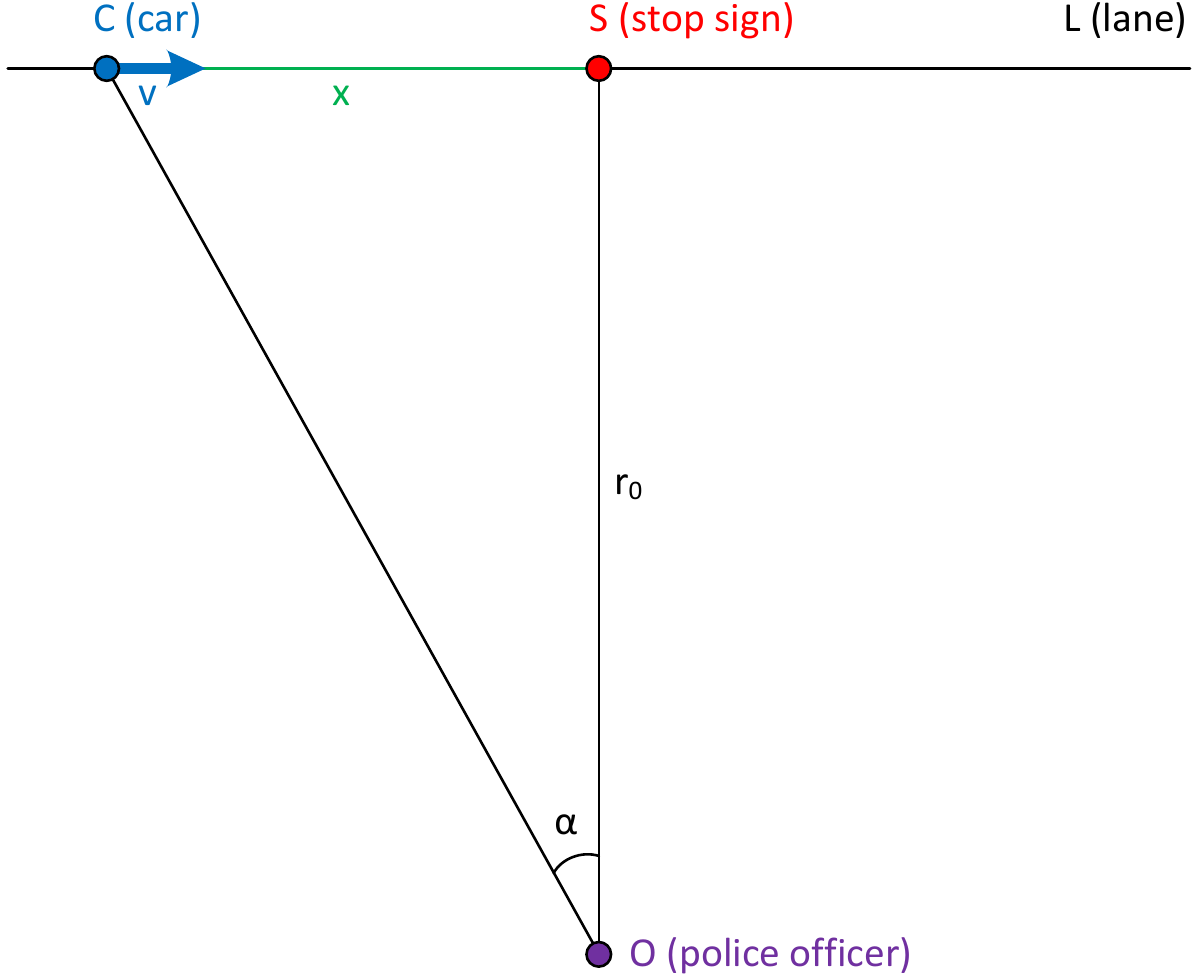}}
\caption{The diagram showing schematically the geometry of the considered case. Car $C$ moves along line $L$. Its current linear speed is~$v$, and the current distance from stop sign~$S$ is~$x$, $|CS|=x$. Another road connects to $L$ perpendicularly at~$S$. Police officer $O$ is located on that road at distance $r_0$ from the intersection, $|OS|=r_0$. The angle between $OC$ and $OS$ is $\alpha$.}
\label{fig:fig1}
\end{figure}

Consider Fig.~\ref{fig:fig1} schematically showing the geometry of the considered case, and assume for a moment that $C$'s linear velocity is constant in time~$t$,
\begin{equation}
v(t) \equiv v_0.
\end{equation}
Without loss of generality we can choose time units $t$ such that $t=0$ corresponds to the moment when $C$ is at~$S$. Then distance~$x$ is simply
\begin{equation}\label{eq:x-linear}
x(t) = v_0t.
\end{equation}

Observer $O$ visually measures not the linear speed of~$C$ but its angular speed given by the first derivative of angle~$\alpha$ with respect to time~$t$,
\begin{equation}\label{eq:omega-def}
\omega(t) = \frac{d\alpha}{dt}.
\end{equation}
To express $\alpha(t)$ in terms of $r_0$ and $x(t)$ we observe from triangle $OCS$ that
\begin{equation}
\tan\alpha(t) = \frac{x(t)}{r_0},
\end{equation}
leading to
\begin{equation}\label{eq:alpha-def}
\alpha(t) = \arctan\frac{x(t)}{r_0}.
\end{equation}
Substituting the last expression into Eq.~(\ref{eq:omega-def}) and using the standard differentiation rules there, i.e., specifically the fact that
\begin{equation}\label{eq:arctan-derivative}
\frac{d}{dt}\arctan f(t) = \frac{1}{1+f^2}\frac{df}{dt},
\end{equation}
where $f(t)$ is any function of $t$, but it is $f(t)=v_0t/r_0$ here, we find that the angular speed of $C$ that $O$ observes as a function of time $t$ is
\begin{equation}
\omega(t) = \frac{v_0/r_0}{1+\left(\frac{v_0}{r_0}\right)^2t^2}.
\end{equation}

\begin{figure}
\centerline{\includegraphics[width=3.5in]{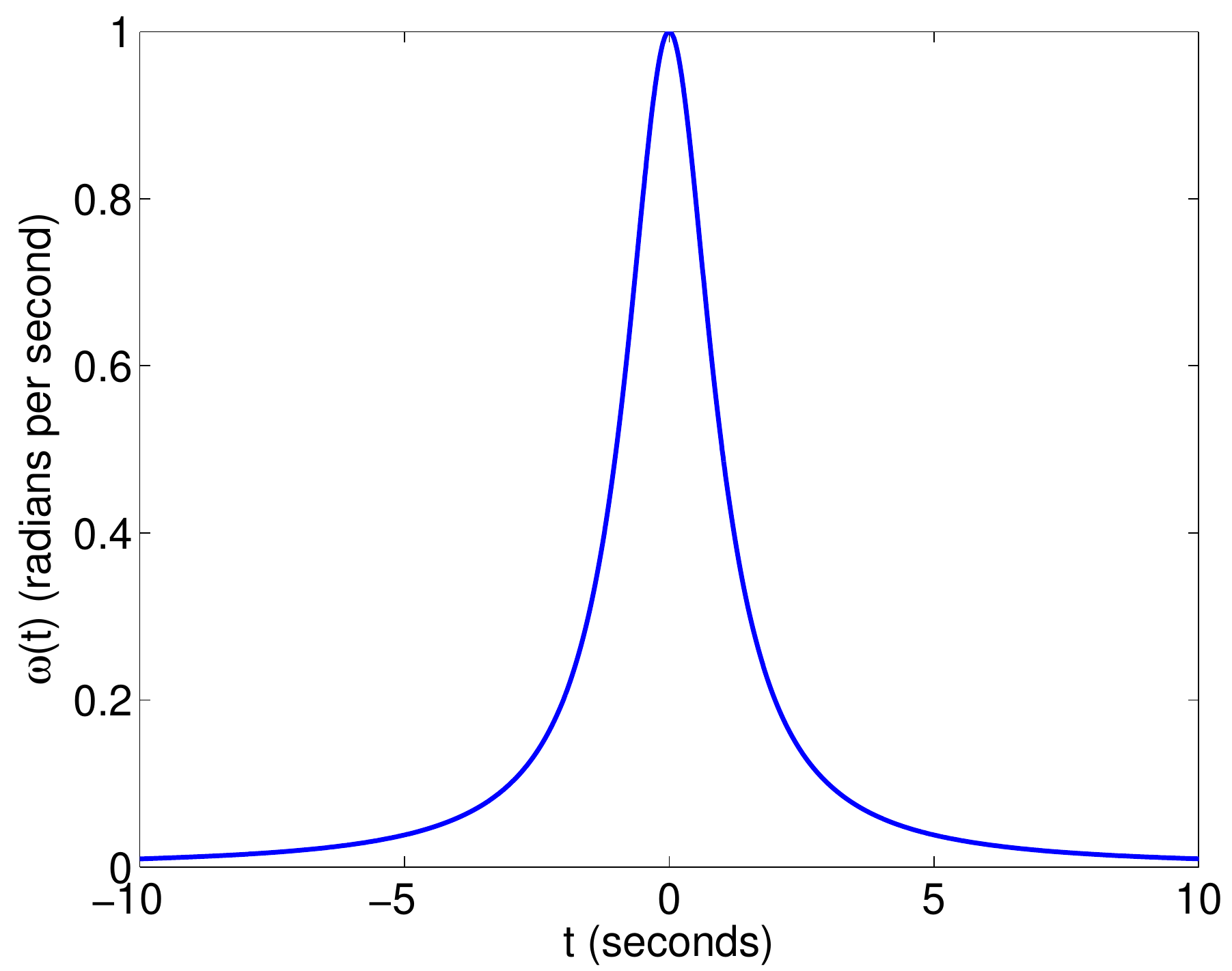}}
\caption{The angular velocity $\omega$ of~$C$ observed by~$O$ as a function of time~$t$ if $C$ moves at constant linear speed $v_0$. The data is shown for $v_0=10\text{ m/s}=22.36\text{ mph}$ and $r_0=10\text{ m} = 32.81\text{ ft}$.}
\label{fig:fig2}
\end{figure}

This function is shown in Fig.~\ref{fig:fig2}. It confirms and quantifies the observation discussed in the previous section, that at $O$, the visual angular speed of $C$ moving at a constant linear speed is not constant. It is the higher, the closer $C$ to $O$, and it goes over a sharp maximum at $t=0$ when $C$ is at the closest point $S$ to $O$ on its linear trajectory $L$.

\section{Constant linear deceleration and acceleration}

In this section we consider the situation closely mimicking what actually happened in the considered case. Specifically, $C$, instead of moving at constant linear speed~$v_0$, first decelerates at constant deceleration~$a_0$, then comes to a complete stop at~$S$, and finally accelerates with the same constant acceleration~$a_0$.

In this case, distance $x(t)$ is no longer given by Eq.~(\ref{eq:x-linear}). It is instead
\begin{equation}\label{eq:x}
x(t) = \frac{1}{2}a_0t^2.
\end{equation}
If this expression does not look familiar, it can be easily derived. Indeed, with constant deceleration/acceleration, the velocity is
\begin{equation}
v(t) = a_0t,
\end{equation}
but by the definition of velocity,
\begin{equation}
v(t) = \frac{dx}{dt},
\end{equation}
so that
\begin{equation}
dx = v(t)\,dt.
\end{equation}
Integrating this equation we obtain
\begin{equation}
x(t) = \int_0^x dx = \int_0^t v(t)\,dt = a_0\int_0^t t\,dt = \frac{1}{2}a_0t^2.
\end{equation}
Substituting the last expression into Eq.~(\ref{eq:alpha-def}) and then differentiating according to Eq.~(\ref{eq:omega-def}) using the rule in Eq.~(\ref{eq:arctan-derivative}) with $f(t) = a_0t^2/(2r_0)$, we obtain the angular velocity of $C$ that $O$ observes
\begin{equation}\label{eq:omega-accel}
\omega(t) = \frac{\frac{a_0}{r_0}t}{1+\frac{1}{4}\left(\frac{a_0}{r_0}\right)^2t^4}.
\end{equation}

\begin{figure}
\centerline{\includegraphics[width=3.5in]{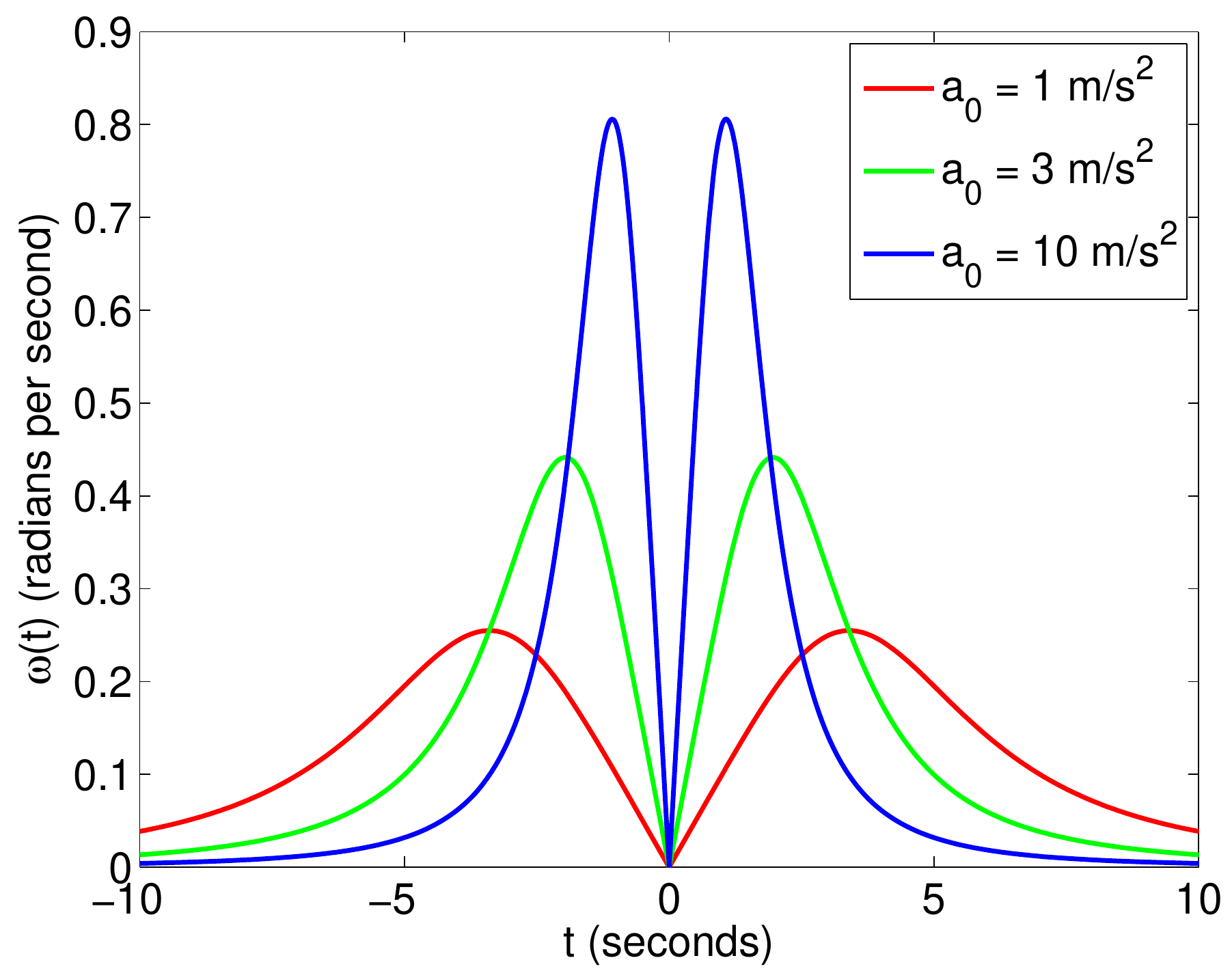}}
\caption{The angular velocity $\omega$ of~$C$ observed by~$O$ as a function of time~$t$ if $C$ moves with constant linear deceleration~$a_0$, comes to a complete stop at~$S$ at time~$t=0$, and then moves with the same constant linear acceleration~$a_0$. The data are shown for $r_0=10\text{ m}$.}
\label{fig:fig3}
\end{figure}

This function is shown in Fig.~\ref{fig:fig3} for different values of~$a_0$. In contrast to Fig.~\ref{fig:fig2}, we observe that the angular velocity of $C$ drops to zero at $t=0$, which is expected because $C$ comes to a complete stop at~$S$ at this time. However, we also observe that the higher the deceleration/acceleration~$a_0$, the more similar the curves in Fig.~\ref{fig:fig3} become to the curve in Fig.~\ref{fig:fig2}. In fact, the blue curve in Fig.~\ref{fig:fig3} is quite similar to the one in Fig.~\ref{fig:fig2}, \emph{except the narrow region\/} between the two peaks in Fig.~\ref{fig:fig3}, where the angular velocity quickly drops down to zero, and then quickly rises up again to the second maximum.

\section{Brief obstruction of view around $t=0$.}

\begin{figure}
\centerline{\includegraphics[width=3.5in]{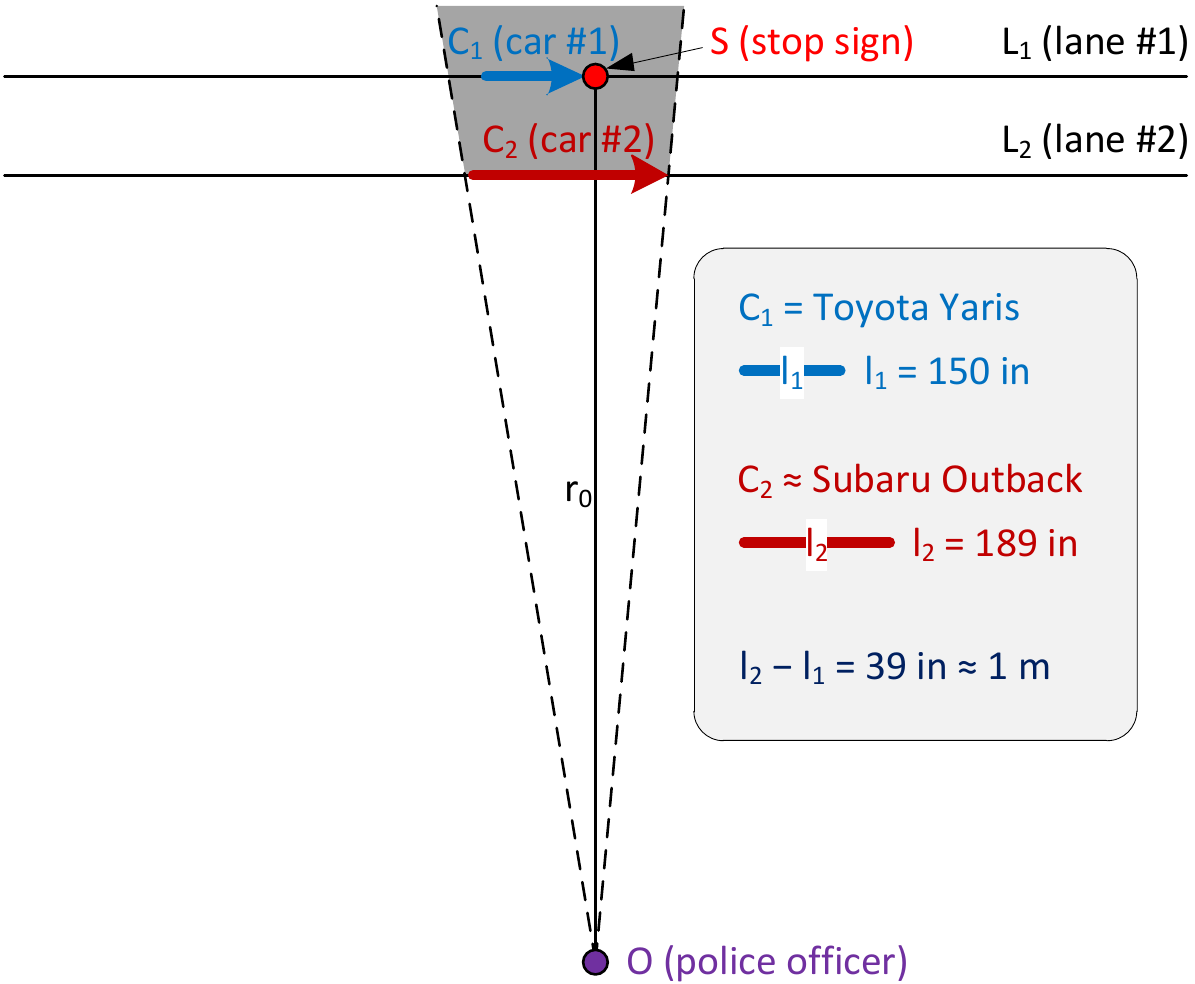}}
\caption{The diagram showing schematically the brief obstruction of view that happened in the considered case. The $O$'s observations of car~$C_1$ moving in lane~$L_1$ are briefly obstructed by another car~$C_2$ moving in lane~$L_2$ when both cars are near stop sign~$S$. The region shaded by the grey color is the area of poor visibility for~$O$.}
\label{fig:fig4}
\end{figure}

Finally, we consider what happens if the $O$'s observations are briefly obstructed by an external object, i.e., another car, see Fig.~\ref{fig:fig4} for the diagram depicting the considered situation. The author/defendant (D.K.) was driving Toyota Yaris (car~$C_1$ in the diagram), which is one of the shortest cars avaialable on the market. Its lengths is $l_1 = 150\text{ in}$. (Perhaps only the Smart Cars are shorter?) The exact model of the other car~($C_2$) is unknown, but it was similar in length to Subaru Outback, whose exact length is $l_2 = 189\text{ in}$.

To estimate times~$t_p$ and~$t_f$ at which the partial and, respectively, full obstructions of view of~$C_1$ by $C_2$ began and ended, we must use Eq.~(\ref{eq:x}) substituting there $x_p = l_2 + l_1 = 8.16\text{ m}$, and $x_f = l_2 - l_1 = 0.99\text{ m}$, respectively. To use Eq.~(\ref{eq:x}) we have to know $C_1$'s deceleration/acceleration $a_0$. Unfortunately, it is difficult to measure deceleration or acceleration without special tools, but we can roughly estimate it as follows. D.K.\ was badly sick with cold on that day. In fact, he was sneezing while approaching the stop sign. As a result he involuntary pushed the brakes very hard. Therefore we can assume that the deceleration was close to maximum possible for a car, which is of the order of $10\text{ m/s$^2$} = 22.36\text{ mph/s}$. We will thus use $a_0 = 10\text{ m/s$^2$}$. Substituting these values of~$a_0$, $x_p$, and $x_f$ into Eq.~(\ref{eq:x}) inverted for~$t$,
\begin{equation}\label{eq:t(x)}
t = \sqrt{\frac{2x}{a_0}},
\end{equation}
we obtain
\begin{eqnarray}
t_p &=& 1.31\text{ s},\label{eq:tp}\\
t_f &=& 0.45\text{ s}.\label{eq:tf}
\end{eqnarray}
The full durations of the partial and full obstructions are then just double these times.

Next, we are interested in time~$t'$ at which the angular speed of $C_1$ observed by $O$ without any obstructions goes over its maxima, as in Fig.~\ref{fig:fig3}. The easiest way to find~$t'$ is to recall that the value of the first derivative of the angular speed at~$t'$ is zero,
\begin{equation}\label{eq:dot-omega-0}
\frac{d\omega}{dt} = \dot{\omega}(t') = 0.
\end{equation}
To find $\dot{\omega}(t)$ we just differentiate Eq.~(\ref{eq:omega-accel}) using the standard differentiation rules, which yield
\begin{equation}
\dot{\omega}(t) = 4\frac{a_0}{r_0} \frac{1-\frac{3}{4}\left(\frac{a_0}{r_0}\right)^2t^4}{\left[1+\frac{1}{4}\left(\frac{a_0}{r_0}\right)^2t^4\right]^2}.
\end{equation}
This function is zero only when the numerator is zero, so that the root of Eq.~(\ref{eq:dot-omega-0}) is
\begin{equation}
t' = \sqrt[4]{\frac{4}{3}}\sqrt{\frac{r_0}{a_0}}.
\end{equation}
Substituting the values of $a_0 = 10\text{ m/s$^2$}$ and $r_0 = 10\text{ m}$ in this expression, we obtain
\begin{equation}\label{eq:t'}
t' = 1.07\text{ s}.
\end{equation}
We thus conclude that time~$t'$ lies between $t_f$ and $t_p$,
\begin{equation}
t_f < t' < t_p,
\end{equation}
and that differences between all these times is actually quite small, compare Eqs.~(\ref{eq:tp},\ref{eq:tf},\ref{eq:t'}).

\begin{figure}
\centerline{\includegraphics[width=3.5in]{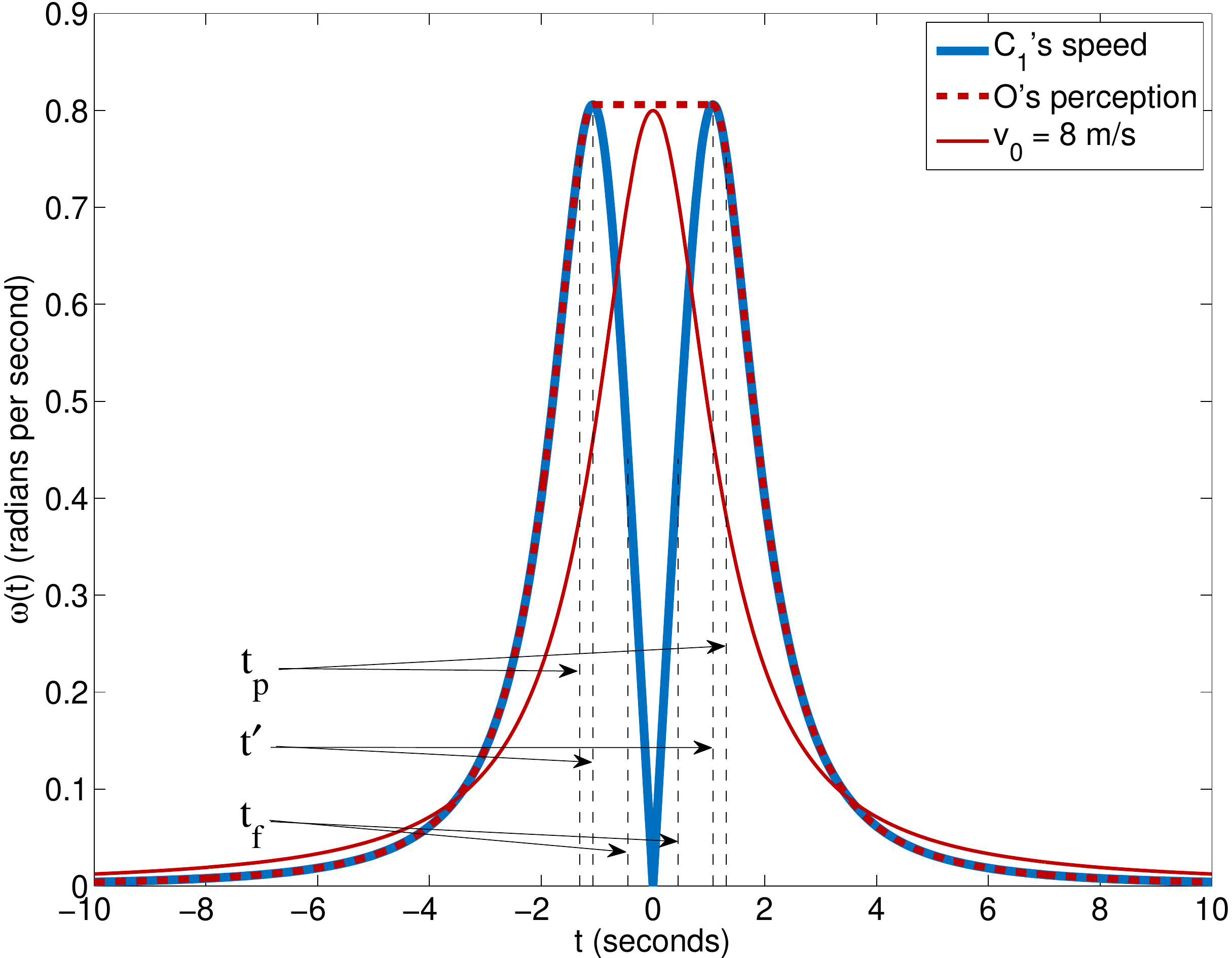}}
\caption{The real angular speed of~$C_1$ is shown by the blue solid curve. The $O$'s interpolation is the dashed red curve. This curve is remarkably similar to the red solid curve, showing the angular speed of a hypothetical object moving at constant linear speed $v_0 = 8\text{ m/s} = 17.90\text{ mph}$.}
\label{fig:fig5}
\end{figure}

These findings mean that the angular speed of $C_1$ as observed by~$O$ went over its maxima when the $O$'s view of~$C_1$ was partially obstructed by~$C_2$, and very close in time to the full obstruction. In lack of complete information, $O$ interpolated the available data, i.e., the data for times $t>t' \approx t_f \approx t_p$, using the simplest and physiologically explainable linear interpolation, i.e., by connecting the boundaries of available data by a linear function. The result of this interpolation is shown by the dashed curve in Fig.~\ref{fig:fig5}. It is remarkably similar to the curve showing the angular speed of a hypothetical object moving at constant speed $v_0 = 8\text{ m/s} \approx 18\text{ mph}$.

\section{Conclusion}

In summary, police officer~$O$ made a mistake, confusing the real spacetime trajectory of car~$C_1$---which moved at approximately constant linear deceleration, came to a complete stop at the stop sign, and then started moving again with the same acceleration, the blue solid line in Fig.~\ref{fig:fig5}---for a trajectory of a hypothetical object moving at approximately constant linear speed without stopping at the stop sign, the red solid line in the same figure. However, this mistake is fully justified, and it was made possible by a combination of the following three factors:
\begin{enumerate}
\item $O$ was not measuring the linear speed of~$C_1$ by any special devices; instead, he was estimating the visual angular speed of~$C_1$;
\item the linear deceleration and acceleration of~$C_1$ were relatively high; and
\item the $O$'s view of~$C_1$ was briefly obstructed by another car~$C_2$ around time~$t=0$.
\end{enumerate}
As a result of this unfortunate coincidence, the $O$'s perception of reality did not properly reflect reality.

\appendix

\section{Two common questions}

\subsection{Is the stop sign fine that high in California?}

The answer is {\it no}. The author did not really know what the fine was since he was not fined. The fine, plus the traffic school (which one wants to take to avoid points on his driving record), is \$287. Therefore the abstract should have read \$300, instead of \$400.

\begin{figure}
\centerline{\includegraphics[width=3.5in]{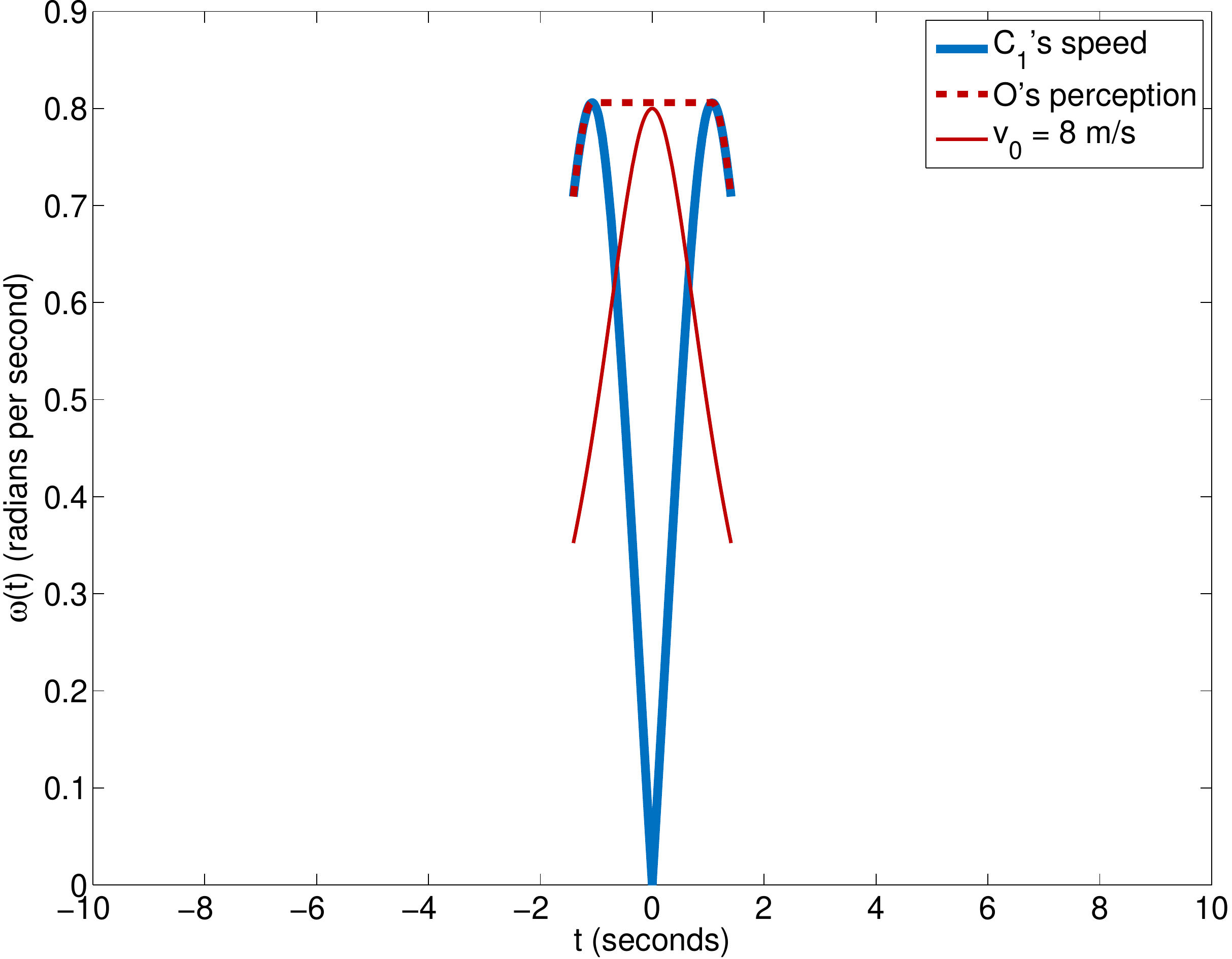}}
\caption{The same data as in Fig.~\ref{fig:fig5} shown for $t\in[-t_b,t_b]$, where $t_b=1.41\text{ s}$.}
\label{fig:fig6}
\end{figure}

\subsection{Are there any flaws in the argument?}

Contrary to common belief, the problem is not that Yaris cannot accelerate that fast. According to the official Toyota specifications, Yaris accelerates to $100\text{ km/h}$ in $15.7\text{ s}$, which translates to $1.77\text{ m/s}^2$. However, this is the {\em average} acceleration, which is not constant. It is well known that most cars accelerate much faster at low speeds than at high speeds, so that the assumption that acceleration $a_0$ was about $10\text{ m/s}^2$ was not unjustified.

This problem of what the exact value of $a_0$ was, becomes actually irrelevant in view of that neither Yaris nor any other car could decelerate or accelerate that fast {\em for that long}, which the author recognized soon after arXival. Indeed, the linear speed of $C_1$ would be too high at $t=\pm10\text{ s}$ in that case. The deceleration/accelaration $a_0\sim10\text{ m/s}^2$ could thus last for only $1$-$2$ seconds.

The question of how the data shown in Fig.~\ref{fig:fig5} would change (presumably not much) if we take into account non-constant $a$ for the whole range of $t\in[-10,10]$, is also irrelevant in view of an additional circumstance brought up by the judge. Both southeast and southwest corners of the intersection in Fig.~\ref{fig:fig4} are occupied by buildings, limiting the view from $O$ to about $l_b=10\text{ m}$ from $S$ along $L_1$. Substituting this $l_b$ instead of $x$ in Eq.~(\ref{eq:t(x)}), we obtain $t_b = 1.41\text{ s}$, which is the time of appearance and disappearance of $C_1$ from $O$'s view obstructed by the buildings. Therefore instead of Fig.~\ref{fig:fig5}, we have Fig.~\ref{fig:fig6}, obtained from Fig.~\ref{fig:fig5} by cutting off all the data outside the range $t\in[-t_b,t_b]$. Clearly, the same conclusions hold, even become stronger.

\end{document}